\documentstyle[epsfig,12pt]{article}
\begin{document}
\newcommand{\gsi}{\,\raisebox{-0.13cm}{$\stackrel{\textstyle>}
{\textstyle\sim}$}\,}
\newcommand{\lsi}{\,\raisebox{-0.13cm}{$\stackrel{\textstyle<}
{\textstyle\sim}$}\,}
\newcommand{\be}{\begin{equation}} \newcommand{\ee}{\end {equation}}

\input epsf
\setlength{\oddsidemargin}{1.2 cm}
\setlength{\topmargin}{1.0 cm}
\setlength{\textwidth}{5.5 true in}
\setlength{\textheight}{8.0 true in}
\setlength{\parskip}{1.5 ex plus0.5ex minus 0.5ex}
\bibliographystyle{unsrt}
\def\question#1{{{\marginpar{\small \sc #1}}}}
\newcommand{\qqbar}{$q \bar{q}~$}
\newcommand{\slsh}{\rlap{$\;\!\!\not$}}     

\rightline{hep-ph/9701290}
\rightline{RAL-97-004}
\baselineskip=18pt
\vskip 0.7in
\begin{center}
{\bf \LARGE Scalar Glueballs and Friends}\\
\vspace*{0.9in}
{\large Frank E. Close}\footnote{\tt{e-mail: fec@v2.rl.ac.uk}} \\ 
\vspace{.1in}
{\it Rutherford Appleton Laboratory}\\
{\it Chilton, Didcot, OX11 0QX, England}\\ 
\end{center}

\begin{abstract}
The $f_0(1500)$ is identified in three glueball favoured production
mechanisms: $p\bar{p}$ annihilation, $\psi \to \gamma f_0$ and Central
Production. The production rate for glueballs in $\psi \to \gamma G$
has been quantified and the $f_0(1500)$ is found to be consistent with
a glueball - $q\bar{q}$ mixture. We illustrate a remarkable property of
central production where kinematic cuts appear to make a systematic
separation between glueballs and $q\bar{q}$ of the same $J^{PC}$. When the
cut favouring glueballs is applied, the $f_0(1500)$ and other enigmatic
states appear prominently while confirmed $q\bar{q}$ states are 
empirically suppressed.
\end{abstract}

\newpage


The lightest glueball in lattice QCD is predicted to be a
scalar\cite{lattice,ct96} for which there is a trio
of suspects 
\begin{itemize}
\item 
$f_0(1500)$\cite{lear,cafe95,bugg}
\item
$f_J(1710)$\cite{weing} where $J=0$ or $2$\cite{pdg96}
\item $f_0(1370)$\cite{pdg96} 
\end{itemize}

Evidence that may help to incriminate one or more of these has emerged
in recent months and in this talk I shall report for the first time
on what may be the long sought
clear dynamical discrimination between $q\bar{q}$ and 
glueballs\cite{ck96,wa102new}.

The new developments relate to the classic production
mechanisms that are believed to favour
glueballs\cite{closerev}:
\begin{enumerate}
\item
Glueballs should be produced in
proton-antiproton annihilation, where the destruction of quarks 
creates opportunity for gluons to be manifested.  This is the Crystal
Barrel \cite{Anis}, and E760 \cite{Hasan1}
production mechanism,
 in which detailed decay systematics of
$f_0(1500)$ have been studied. The empirical situation with regard to
$f_J(1710)$ is currently under investigation. 
\item
Glueballs should be enhanced compared to ordinary mesons in radiative
quarkonium decay. The $f_0(1500)$ and $f_J(1710)$
are produced
in radiative $J/\psi$ decay at a level typically of $\sim1$ part per
thousand. 
\item
Glueballs are hypothesised to be favoured over ordinary mesons in the
central region of high energy scattering processes, away from beam and
target quarks.  The $f_J(1710)$ and possibly the $f_0(1500)$ have been
seen in the central region in $pp$
collisions\cite{Kirk,Gentral}.
\end{enumerate}

In recent months there have been breakthroughs in each of these three
areas, in particular that the $f_0(1500)$ is now definitely seen
in central production\cite{ck96,wa102new}. I shall deal with each in turn.

\section{$p\bar{p}$ annihilation}

The first clear sightings of the $f_0(1500)$ were in $p\bar{p}$
annihilation and their implications for glueball phenomenology have
been analysed in some detail\cite{cafe95}. This is widely regarded 
as the first potential clear sighting of a glueball mixed in with
other scalar $q\bar{q}$ states\cite{leap,landua96}. The completion
of the programme in $p\bar{p}$ annihilation is now governed
principally by the closure of LEAR at the end of 1996. Analysis of
the data will continue so that during the next three years we
may obtain information about the decay channels for this state together with
confirmation of other scalar states in this
mass region, such as $a_0(1450)$. 

This is for the future and I shall not repeat here my
recent summary which may be found in ref.\cite{leap}. 
I want instead to concentrate on the new developments in the other two
favoured mechanisms.

\section{$J/\psi \to \gamma f_J$}

There have been significant advances in the quantitative analysis of
glueball production in the second of these processes, $J/\psi \to \gamma f_J$
\cite{cfl96,cak}.

The production rate for glueballs or $q\bar{q}$ in $J/\psi \to \gamma R$
is quantified in ref\cite{cfl96}; specifically, 
for scalar mesons 
\begin{equation} \label{0++}
10^3  br(J/\psi \rightarrow \gamma 0^{++}) = (\frac{m}{1.5\; {\rm GeV}})  
(\frac{\Gamma_{R\rightarrow gg}}{96\; {\rm MeV}})  \frac{x|H_S(x)|^2}{35}.
\end{equation}
This is to be compared with the analogous formula for a tensor meson:
\begin{equation} \label{2++}
10^3  br(J/\psi \rightarrow \gamma 2^{++}) = (\frac{m}{1.5\; {\rm GeV}})  
(\frac{\Gamma_{R\rightarrow gg}}{26\; {\rm MeV}})  \frac{x|H_T(x)|^2}{34}.
\end{equation}
where $x|H_J|(x)|^2$ is a loop integration in pQCD whose magnitude
is shown in fig. 1. 
Having scaled the expressions this way, because $\frac{x|H_J|^2}{30-45}
\sim O(1)$ in the $x$ range relevant for production of 1.3 - 2.2 GeV
states,  we see immediately that the magnitudes 
of the branching  ratios are driven by the denominators 96 and 26 MeV
for $0^{++}$ and $2^{++}$. Thus if
a state $R_J$ is produced in $J/\psi \rightarrow \gamma X$ at
$O(10^{-3})$ then $\Gamma (R_J \rightarrow gg)$ will typically be of  
the order $100$ MeV for $ 0^{++}$, $O(25 ~ {\rm MeV})$ for $2^{++}$.

This immediately shows why the $2^{++}$ $q \bar{q}~$ states are prominent: A
$2^{++}$ state with a total width of $O(100 ~\rm{MeV})$ (typical for
$2^{++}$ $q \bar{q}~$'s in this mass range\cite{cafe95,barnes95}) will be
easily visible in $J/\psi \rightarrow \gamma 2^{++}$ with branching
fraction $O(10^{-3})$, while remaining consistent with
\begin{equation} \label{f320} 
br(R[Q\bar{Q}] \rightarrow gg) = 0(\alpha^2_s) \simeq 0.1-0.2.
\end{equation}
For a glueball, on the other hand, one expects\cite{cfl96,cak}
\begin{equation} \label{f320} 
br(R[G] \rightarrow gg) = 0(1).
\end{equation}

From our relations above, we see that for a $0^{++}$ to be produced at
the $10^{-3}$ level in $J/\psi$ radiative decay it must either have a
large gluonic content and width $O(100)$ MeV or, if it is a $q \bar{q}~$
meson, it must have a very large width,
 $\,\raisebox{-0.13cm}{$\stackrel{\textstyle>}{\textstyle\sim}$}\, 500$ MeV. 

When applied to the data  the conclusions are:

(i) The $f_0(1500)$ is probably produced at a rate too high to be a
$q\bar{q}$ state.  The average of world data suggests it is a
glueball-$q \bar{q}$ mixture.  

(ii) The $f_J(1710)$ is produced at a rate which is consistent with
it being $q\bar{q}$, only if $J=2$.  If $J=0$, its production
rate is too high for it to be a pure $q\bar{q}$ state but is consistent
with it being a glueball or mixed $q \bar{q}$-glueball having a large
glueball component. 

(iii) The $f_0(1370)$ may have some glueball admixture in its wavefunction
but contains $q\bar{q}$ dominantly.

These studies were stimulated by the advances in lattice QCD which 
predict that the lightest
``ideal" (i.e., quenched approximation) glueball be $0^{++}$, with
state-of-the-art mass prediction of $1.61 \pm 0.07 \pm 0.13$ GeV
(where the first error is statistical and the second is systematic)\cite{ct96}.
This subsumes the values in the literature of $1.55 \pm 0.05$ GeV\cite{ukqcd}
and $1.74 \pm 0.07$ GeV\cite{weing}.  That lattice QCD is now
concerned with such fine details represents considerable advance in
the field and raises both opportunity and enigmas. It
encourages serious consideration of the further lattice predictions
that the $2^{++}$ glueball lie in the $2$ GeV region as well
as implying that scalar
mesons in the $1.5-1.7$ GeV region merit special attention.  Amsler
and Close\cite{cafe95} initially pointed out that the $f_0(1500)$ shares
features expected for a glueball that is mixed with the nearby
isoscalar members of the $^3P_0$ $q\bar{q}$ nonet.  If the $f_J(1710)$
proves to have $J=2$, then it is not a candidate for the ground state
glueball and the $f_0(1500)$ will be essentially unchallenged.
On the other hand, if the $f_J(1710)$ has $J=0$ it becomes a potentially
interesting glueball candidate.  Indeed, Sexton, Vaccarino and
Weingarten\cite{weinprl} argue that $f_{J=0}(1710)$ should be
identified with the ground state glueball, based on its similarity in
mass and decay properties to the state seen in their lattice
simulation.  

The properties of the $f_J(1710)$, and the production of this state
together with $f_0(1500)$ in the three glueball favoured processes, are
now the leading questions for experiment.

\section{A Glueball-$q\bar{q}$ filter in Central Hadron Production}

``Central production" where
the produced mesons have no memory of the
flavour of the initiating hadrons\cite{closerev}
 and are excited via the gluonic fields of
the ``Pomeron"\cite{landshoff} is the final example in our trinity of
glueball production mechanisms. However,
although at first sight this is an environment where
the production of 
glueballs may be especially favoured, the reality is more complicated.

First there is the well known problem that non diffractive transfer
of flavour (Regge exchange) can
contaminate this simple picture and lead to the appearance of $q\bar{q}$
mesons in the central region. Furthermore, even for the diffractive production,
momentum transfer between the gluons of the Pomeron and 
the aligned constituents of the produced meson may
lead to either $gg$ or $q\bar{q}$ states. 
The former may be
favoured relative to $q\bar{q}$ production due to colour factors
 but unless further cuts are made to enhance the $gg$ signal, the appearance of
 novel states in central production is not of itself definitive evidence
for a glueball. This is clear from the data which show well known $q\bar{q}$
states, such as $f_1(1285)$, alongside potential glue states (as, for example,
in the $4 \pi$ mass spectrum, ref.\cite{interfere,wa9196}).
 Similar mixtures were seen in other
channels, for example $\pi \pi$ where there is not only a clear $f_0(980)$
but also the $f_2(1270)$ and even $\rho(770)$. This confusing menagerie 
of states, with assorted $J^{PC}$ and with established $q\bar{q}$ alongside
``interesting"  states, has caused central production to be an enigma.

The first clue on how to decode the central production came
with the empirical observation earlier in 1996\cite{wa9196} 
that the
structures seen in central production are a function of the topology,
and depend on whether events are classified as either $LL$ or $LR$
(``left left" or ``left right" in the sense of how the beams scatter
into the final state relative to the initial direction). Specifically, when
the two beams scatter into opposing hemispheres ($LR$ as defined in
ref.\cite{wa9196}) the $f_1(1285)~ ^3P_1$ 
$q\bar{q}$ is clearly visible (fig 2a) 
whereas in the same side configuration 
($LL$) it is less prominent relative to the structures in the
$1.4 - 2$~GeV mass range. (fig 2b).
Such discrimination has also been seen for the $f_2(1270)$ 
$^3P_2$ $q\bar{q}$ relative to the enigmatic $f_0(980)$ in the $\pi \pi$
channel\cite{wa9196}. Similar phenomena have recently been noted also
in Fermilab data \cite{fermi} and, in retrospect, at the ISR\cite{isr}.

This phenomenon led us to reconsider the  mechanisms for
the production of $gg$ and $q\bar{q}$ in the central region\cite{ck96}.
Our notation is that in the centre of mass frame
the initial protons have $p =(P+M^2/2P;p_T=0,p_L =P),
 ~ q =(P+M^2/2P;q_T = 0,q_L =-P)$,
the outgoing protons having
respectively momenta $p' \equiv (p'_L = x_a p; ~ p'_T)$ 
and $q' \equiv (q'_L = x_b q; ~ q'_T)$ 
and $x_F \equiv x_a - x_b$. 
The data have historically been
presented as a function of $M_R^2 \sim (1-x_a)(1-x_b)s$ with some
separation as a function of 
$t_a \sim -\frac{({p'_T})^2}{1-x_a}; 
~ t_b \sim -\frac{({q'_T})^2}{1-x_b}$. The longitudinal momenta,
$p_L', ~ q_L'$ therefore control the $M_R$ distribution.
The topological separation into $LL$
and $LR$ was novel and independent of the magnitudes of $t_{a,b}$\cite{wa9196}.
We  suggested that it is driven primarily by the variable
$dP_T \equiv |\vec{p'_T} - \vec{q'_T}|$ and that
$gg$ configurations are enhanced in kinematic
configurations where the gluons can flow ``directly" into the final state
with only small momentum transfer, in particular when $dP_T \to 0$ (driven
by fig. 3b rather than 3a).
This configuration may also enhance $0^{++}, ~ 2^{++}$ mesons made from
bosons in $S$-wave relative to their $q\bar{q}$ $^3P_{0,2}$ counterparts.

When the $\vec{p'_T}$ and
$\vec{q'_T}$ are co-moving and of equal magnitude such that $dP_T \to 0$, 
they tend to produce an overall transverse boost 
of the meson $R$ but with limited relative (internal) momentum: the
resulting configuration for 
$R$ will be strongly coupled to an $S$-state. 
By contrast, when the $\vec{p'_T}$ 
and $\vec{q'_T}$ are equal in magnitude but
anti-aligned (so $dP_T$ is large) then 
there can be significant relative $l_T$ ($\sim dP_T$) within $R$;
 excitation of the $l_T$ degree of freedom corresponds to 
$P$-wave (and higher orbitals) in the static limit.
Thus by making the selection on data
that $dP_T \to 0$, there is the possibility that $q\bar{q}$ with 
$J^{PC} = 0^{++},1^{++},2^{++}$ (which are all $P$-wave composites) will
be suppressed relative to glueballs (or at least, 
relative to $S$-wave bound states of bosons with these $J^{PC}$).

 The above intuitive picture was only a first sketch of the
full dynamics at best, designed principally as a starting point that is
qualitatively consistent with the
pattern of the $LL$ and $LR$ topologies. Nonetheless, when
the WA102
collaboration at CERN tested the suggestion that
the variable $dP_T \equiv |\vec{p'_T} - \vec{q'_T}|$
controls the dynamics, they found a remarkable picture.

The  $dP_T \geq 0.5$ GeV (fig 3a) is similar to the original $LR$
sample as expected. The sample with $0.2~ GeV \leq dP_T \leq 0.5 ~
GeV$ (fig 3b) shows the $f_1(1285)$ becoming suppressed and a sharpening of the
$f_0(1500)$ and $f_{2}(1900)$ structures. However, the most
dramatic effect is seen in the  $dP_T \leq 0.2~GeV$ sample (fig 3c) where the
$f_1(1285)$, a $q\bar{q}$ state, has essentially disappeared while the 
$f_0(1500)$ and $f_{2}(1900)$ structures have become more clear. 
These surviving structures have been identified as 
glueball candidates: the $f_0(1500)$ is motivated by lattice QCD while
the $f_2(1900)$ is noted to have the right mass to lie on the Pomeron
trajectory\cite{pvl}.
The $f_0(1500)$ is rather clean and
appears at $dP_T \to 0$ with a 
shape and mass that are not inconsistent with what is seen in
$p\bar{p}$ annihilation.
This is in contrast to the full data sample of the present
experiment where this state interfered with the $f_0(1370)$ and was
shifted to a lower mass ($\sim 1440$ MeV)\cite{interfere} and
with a much narrower width ($\sim 60$ MeV).

Similar cuts have been applied to the $\pi\pi$, $K\bar{K}$ $\eta \pi \pi$
and $K\bar{K} \pi$
data \cite{wa102new}).

For the $\pi^+ \pi^-$ mass spectrum for $dP_T < 0.2$ GeV there is
effectively no $\rho$ or $f_2(1270)$ signals. These only become apparent
as $dP_T$ increases. However, the $f_0(980)$, which is responsible
for the sharp drop in the spectrum around 1 GeV, is clearly visible in the
small $dP_T$ sample, (figures 4)

Figures (5) show the effect of the $dP_T$ cut on the $K^+K^-$ spectrum. The
$f_2(1525)$ is produced dominantly at high $dP_T$ whereas the $f_J(1710)$
is produced dominantly at low $dP_T$.
In the channels $K\bar{K} \pi$
and $\eta \pi \pi$
the $f_1(1285)$ and $f_1(1420)$ are more prominent when $dP_T > 0.5 ~ GeV$
and start to disappear at low $dP_T$.

It would appear that the undisputed $q\bar{q}$ states (i.e. $\rho$, $f_2(1270)$,
$f_1(1285)$, $f_2(1525)$) are suppressed as $dP_T$ goes to zero whereas the
glueball candidates $f_J(1710)$, $f_0(1500)$ and $f_2(1930)$ survive. It is
also intersting to note that the enigmatic $f_1(1420)$ disappears
at low $dP_T$ while the $f_0(980)$, a state that possibly has a strong 
admixture of $K\bar{K}$ in its wavefunction\cite{wein,torn},does not.

Thus we have a tantalising situation in central production of mesons.
We have stumbled upon a remarkable empirical feature 
that does not appear to have been noticed previously. Although
its extraction via the $dP_T$ cut was inspired by intuitive arguments
following the observation of an $LL-LR$ asymmetry, we have no
simple dynamical explanation. Nonetheless, the empirical message is
dramatic enough to stand alone and thereby
we are suggesting\cite{ck96} that a systematic study of meson production as
a function of $dP_T \equiv |\vec{p'_T} - \vec{q'_T}|$ holds special
promise for isolating the systematics of meson production in the central region
and in filtering $q\bar{q}$ mesons from those with significant $boson$-$boson$
content. The latter include $K\bar{K}$ molecular bound states (or $s\bar{s}$
states with significant $K\bar{K}$ component in the wavefunction), $pomeron$
- $pomeron$ states and $glueballs$.
Our selection
procedure will need to be tested  further in future experiments in
order to determine  the extent of its empirical validity. In turn we
hope  thereby that its dynamical
foundations may be put on a sounder footing and the filtering of glueballs be
made a practical reality.

\vskip 0.4cm
\hspace*{2em}
FEC is partially
supported by the European Community Human Mobility Program Eurodafne,
Contract CHRX-CT92-0026. I am grateful to my collaborators, C.Amsler,
G.Farrar, Z.P.Li and A.Kirk and also
to the WA102 collaboration who were involved with me in various aspects
of the work reported here.

\newpage

\begin{figure}
\epsfxsize=\hsize
\epsffile{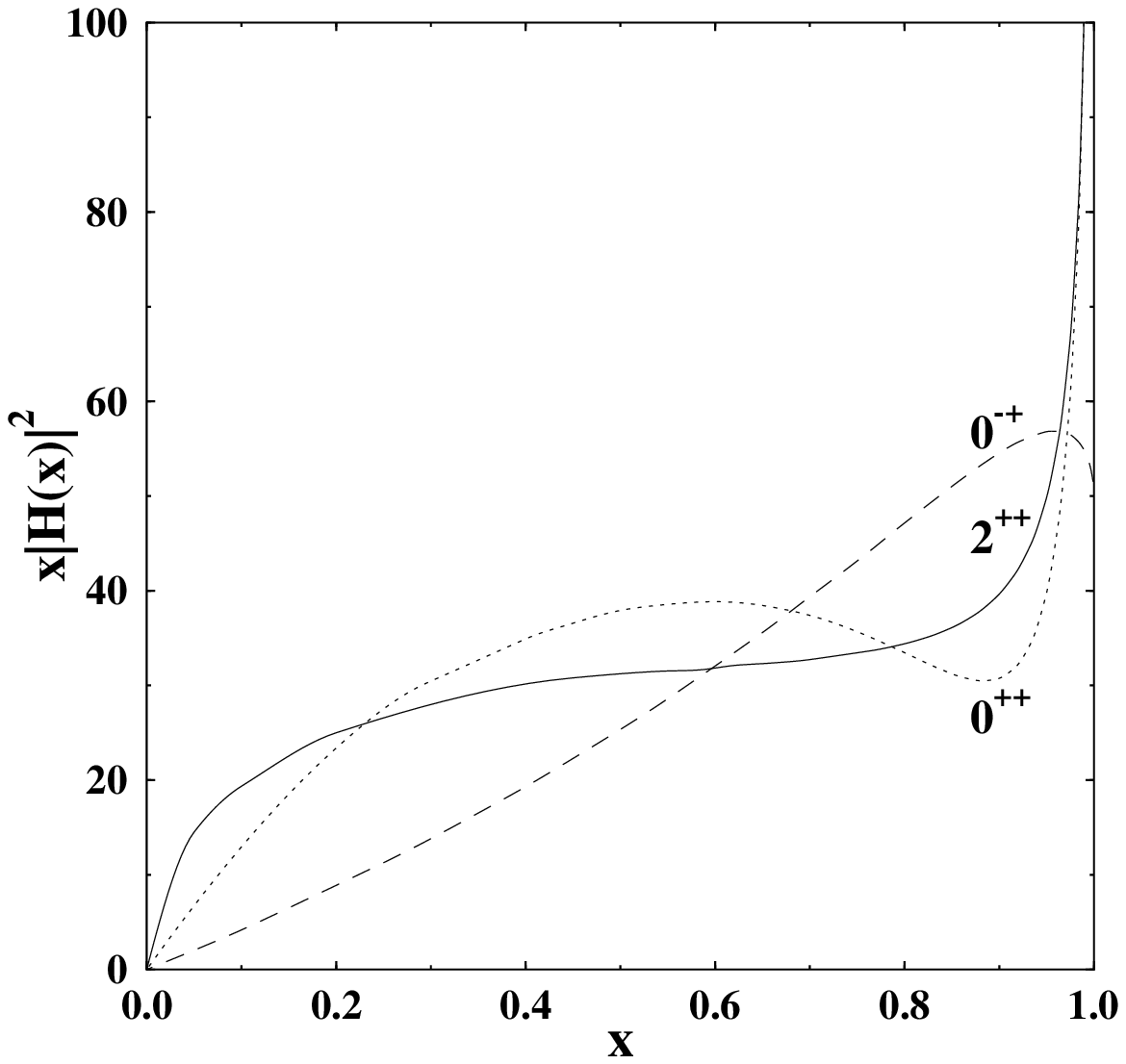}
\caption{Computation of $\psi \to \gamma gg \to \gamma R$ in pQCD
involves a loop integral. The magnitude of this loop integral, $x|H|^2$ 
is plotted versus $x$ for $0^{++}$ 
(dotted), $0^{-+}$ (dashed) and $2^{++}$ (solid); $x=1-(\frac{m_R}{m_V})^2$.}  
\label{HJ}
\end{figure}

\begin{figure}
\epsfxsize=\hsize
\epsffile{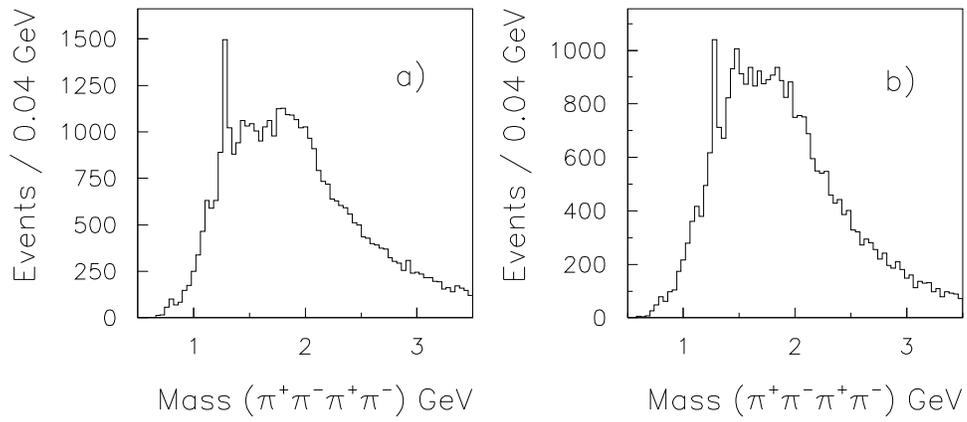}
\caption{The 4$\pi$ mass spectra from WA91 for (a) LR and (b) LL
topologies.}  
\label{lllr}
\end{figure}
\begin{figure}
\epsfxsize=\hsize
\epsffile{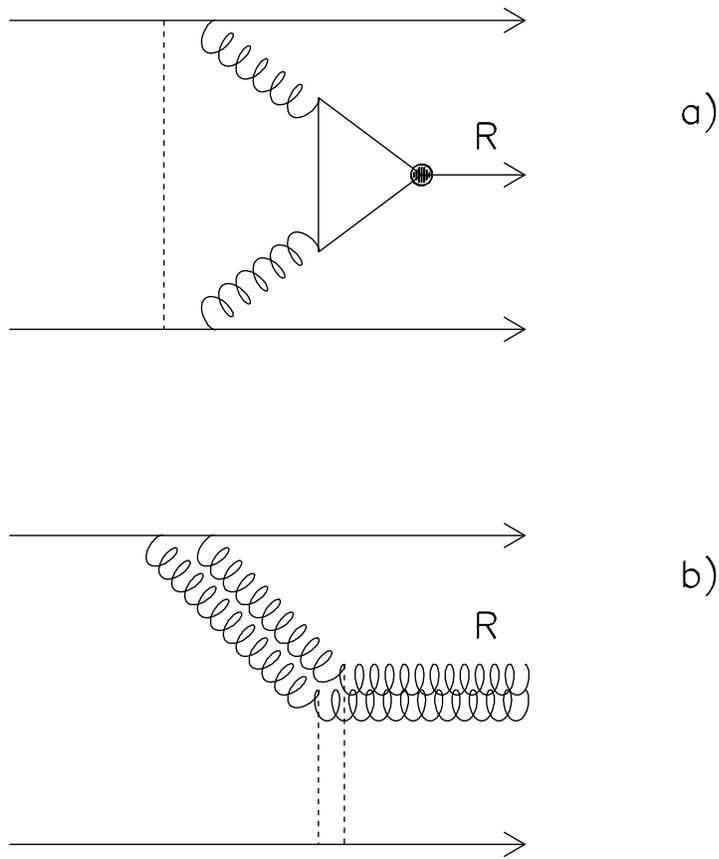}
\caption{ (a) Two gluons with large $p_L$ fuse to make a meson $R$. 
This requres considerable rescattering to produce aligned constituents
in the exclusive production of a meson $R$.
(b) Diffractive scattering of a gluonic Pomeron to produce a glueball.
By analogy with the diffractive photoproduction of $^3S_1$ vector mesons,
this can be favourable to $0^{++}$ and $2^{++}$ ``S"-wave glueballs.}
\label{fi:1}
\end{figure}
\begin{figure}
\epsfxsize=\hsize
\epsffile{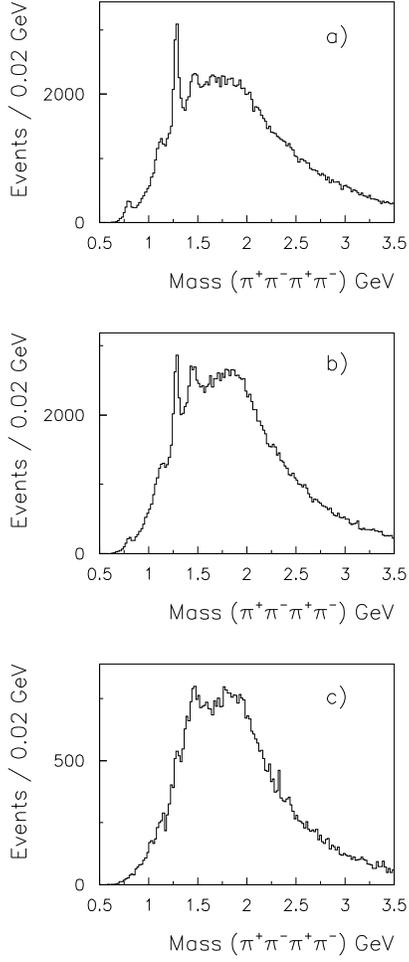}
\caption{The 4$\pi$ mass spectra (a) With $dP_T > 0.5$~GeV
exhibiting a clear $f_1(1285)$; (b) $0.2 < dP_T < 0.5$~GeV
(c) $dP_T < 0.2$~GeV where the $f_1(1285)$ has disappeared while
the $f_0(1500)$ is seen more clearly.}
\label{4pi}
\end{figure}
\begin{figure}
\epsfxsize=\hsize
\epsffile{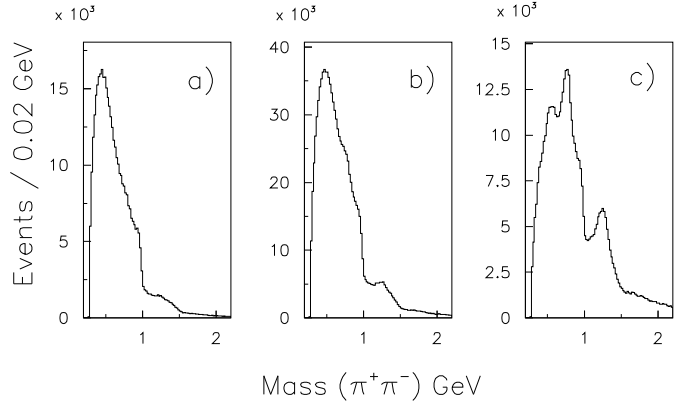}
\caption{The $\pi\pi$ mass spectrum for 
a) $dP_T<0.2$ GeV, b) $0.2 <dP_T <   0.5$ GeV and c) $dP_T >   0.5$ GeV.
The $\rho(770)$ and $f_2(1270)$ only become apparent as $dP_T$ increases.
The $f_0(980)$ is responsible for the sharp drop in the spectrum round 1 GeV
and is clearly visible in the low $dP_T$ sample.}
\label{fi:2pi}
\end{figure}
\begin{figure}
\epsfxsize=\hsize
\epsffile{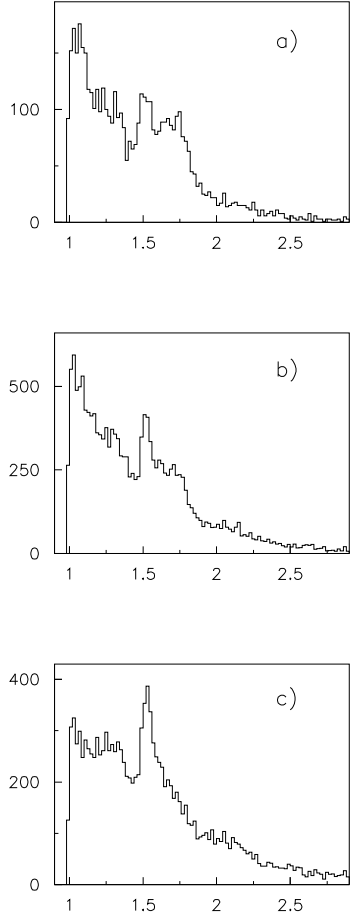}
\caption{$K^+K^-$ mass spectrum for a) $dP_T <   0.2$ GeV, 
b) $0.2 <   dP_T <   0.5$ GeV and c) $dP_T >   0.5$ GeV.The $f_2(1525)$ is
produced dominantly at high $dP_T$, fig (c), whereas the $f_J(1710)$ is
produced dominantly at low $dP_T$, fig (a). }
\label{fi:2k}
\end{figure}

\end{document}